\documentclass[12pt,a4paper]{article}
  \usepackage{epsfig}
  \newcommand{\be}{\begin{equation}}
  \newcommand{\ee}{\end{equation}}
  \newcommand{\bea}{\begin{eqnarray}}
  
  \newcommand{\eea}{\end{eqnarray}}

  \newcommand{\spur}[1]{\not\! #1 \,}
  \def\slash#1{\setbox0=\hbox{$#1$}#1\hskip-\wd0\dimen0=5pt\advance
  \dimen0 by-\ht0\advance\dimen0 by\dp0\lower0.5\dimen0\hbox
  to\wd0{\hss\sl/\/\hss}}
  \def\Black{}
  \def\Blue{}
  \def\Brown{}
  
\begin{document}
  \begin{titlepage}
  \title{\hfill $\mbox{\small{\begin{tabular}{r} $\Blue{\rm
  BA-TH/422-01}$\\ $\Blue{\rm  Napoli-DSF~2001/27}$
  \end{tabular}}}$ \\[1truecm]
  \Brown Charming penguin contributions to charmless $B$ decays into two pseudoscalar
  mesons}
  \vspace{+1truecm}
  \author{\Black C. Isola$^{a}$, M. Ladisa$^{b}$, G. Nardulli$^{c}$, T. N. Pham$^{a}$,
  P. Santorelli$^{d}$}
  \date{~}
  \maketitle
  \vspace{-2truecm}
  \begin{it}
  \begin{center}
$^a$Centre de Physique Th{\'e}orique, \\
Centre National de la Recherche Scientifique, UMR 7644, \\
{\'E}cole Polytechnique, 91128 Palaiseau Cedex, France\\
\vspace*{0.3cm}
$^b$Physics Department, Technion-Israel Institute of Technology, Haifa 32000, Israel\\
\vspace*{0.3cm}
$^c$Dipartimento di Fisica dell'Universit\`{a} di Bari, Italy\\
Istituto Nazionale di Fisica Nucleare, Sezione di Bari,Italy\\
\vspace*{0.3cm} $^d$Dipartimento di Scienze Fisiche,
Universit\`{a} di Napoli "Federico II", Italy\\Istituto Nazionale
di Fisica Nucleare, Sezione di Napoli, Italy \vspace*{0.3cm}
  \end{center}
  \end{it}

\begin{abstract}
\noindent We present estimates of the charming penguin
contribution to $B \to K\pi,\pi\pi,K\eta,K\eta^{\prime}$ decays
due to intermediate charmed meson states. We find that this
contribution is indeed significant for $B\to K\pi$ decays, and its
inclusion, together with the tree and penguin terms,  produces
large branching ratios in agreement with data, though the
analysis is affected by large theoretical uncertainties. On the
other hand, for $B \to \pi\pi,\,K\eta,\, K\eta^\prime$ decays, the
effect of the charming penguin contribution is more modest. We also compute CP
asymmetries for  $B \to K\pi,\,\pi\pi$ decays and we obtain rather
large results.
\end{abstract}

\vspace{2.5truecm}
\thispagestyle{empty}
\end{titlepage}
\setcounter{page}{2}
\newpage

\section{Introduction}
\noindent In a recent paper \cite{noi1}, hereafter referred to as
${\cal I }$, we gave an estimate of the so-called charming penguin
\cite{Ciuchini1}, \cite{strikeback} contributions to the decays $B
\to K \pi$. This is a long-distance part of the decay amplitude
whose imaginary part results from the decay chains
\begin{eqnarray}
  B &\to& D(D_s)\to K\pi\ ,\cr
  B&\to & D^{*}(D_s^{*})\to K\pi\ ,\label{cp}
\end{eqnarray}while the real part can be computed by a tree diagram
 of the effective chiral lagrangian for heavy mesons
\cite{wise}-\cite{Deandrea2}. In the present paper, we shall call
this amplitude $A_{ChP}$.
The
relevance of these contributions for $B\to K\pi$ decays was first
pointed out in \cite{Nardulli}; though suppressed in the
factorization approximation, these terms are enhanced by the
Cabibbo-Kobayashi-Maskawa (CKM) matrix factor $V_{cb}V_{cs}^*$ in
comparison to the short distance terms, {\it i.e.}, contributions
arising from the Tree and Penguin terms in the factorization
approximation, whose amplitude we call here $A_{T+P}$. In ${\cal I
}$ we have shown that, even taking into account the uncertainties
inherent to this calculation, the contribution of the charming
penguins contributions to the decay channels $B^+ \to K^0 \pi ^+$ and $B^0 \to
K^+ \pi ^-$ is indeed significant and can explain the difference
between the data and the result obtained by $A_{T+P}$. In the
present paper, we wish to extend the analysis to cover other $B$
decay channels with a $K\pi$ pair in the final state, as well as
other charmless $B$ decays into two pseudoscalar mesons, {\it
i.e.},
\begin{eqnarray} B&\to &  \pi\pi\ ,\label{pipi}\\
  B &\to& K\eta \ , \label{keta}\\
  B &\to&  K\eta^\prime \ .\label{ketaprimo}
\end{eqnarray}
For the processes (\ref{keta}) and (\ref{ketaprimo}) we add to
$A_{T+P}$ the the charming penguin contributions with $
D^{(*)},\,D_s^{(*)}$ intermediate charmed meson states; for the
$\pi\pi$ final state, the charming penguin contribution is obtained
by $ D^{(*)},\,D^{(*)}$ intermediate states. All the relevant
formulae are presented in ${\cal I }$ and can be applied  here
with some obvious changes. For example the substitution $K\to\pi$
for the channel (\ref{pipi}) or the substitution of the pion
physical constants with the analogous observables of $\eta$ and
$\eta^\prime$ for the channels (\ref{keta}) and (\ref{ketaprimo}).
A few points, however, deserve a more detailed discussion; let us
examine them in the next section.

\section{Discussion on the method and its uncertainties}

The procedure for obtaining the real  part is based on the use of
an effective field theory satisfying chiral symmetry as well as
heavy flavor symmetries. The main point in this procedure is the
following approximation (we take $B^+\to K^0\pi^+$ as the
representative channel for the $B\to PP$ decays):
\begin{eqnarray}
  A_{ChP}&=&{G_F \over \sqrt{2}} \, a_2\, V_{cb}^* V_{cs}<K^0\pi^+|:J_{\mu}(0)\hat
  J^{\mu}(0):|B^+>\approx\cr&\approx&{G_F \over \sqrt{2}}\, a_2\,V_{cb}^* V_{cs}
  \int\frac{d\vec n}{4\pi}<K^0\pi^+|T\left\{J_{\mu}(x_0)\hat
  J^{\mu}(0)\right\}|B^+>\label{eq:1}
\end{eqnarray}
with $J_{\mu} = \bar{b}\gamma_{\mu}(1 - \gamma _5)c$ and $\hat
J_{\mu} = \bar{c}\gamma_{\nu}(1 - \gamma _5)s$; moreover $a_2 =(c_2+c_1/3)\;\; (=1.03)$ where
$C_1$ and $C_2$ are Wilson coefficients, and
\begin{equation}
x_0^\lambda\,=\,\left(0,\,{\vec n}/{\mu}\right)\ ,
\end{equation}
where $|\vec n|=1$ and $\mu$ is a scale representing the onset of
the scaling behaviour. 
In (\ref{eq:1}) we have not considered color octet
operators which would have given no contribution as we consider only the
color-singlet physical intermediate states.
This approximation is based on the light-cone expansion \cite{wilson},
\cite{brandt}, which in the present case reads:
\begin{equation}
<K^0\pi^+|T\left\{J_{\mu}(x_0)\hat
  J^{\mu}(0)\right\}|B^+>\,\approx\,<K^0\pi^+|:J_{\mu}(0)\hat
  J^{\mu}(0):|B^+>\,+\,{\cal O}(x_0^2)\, ,
  \label{eq:expansion}
\end{equation}
where the ${\cal O}(x_0^2)$ terms are negligible for $\mu=1/|x_0|$
sufficiently large ($\mu\sim m_b$); clearly the integral over
$\vec n$ corresponds to an average over the directions of $\vec
x_0$.
Moreover, the non trivial scale dependence of the r.h.s.  of eq.
(\ref{eq:expansion}) is matched by the Wilson coefficient,
to give scale-independent physical observables. Also the l.h.s.
contains a scale (the cut-off). Ideally this scale dependence should be
cancelled by the short distance coefficient as well; in practice, however,
the cancellation is not complete as we make a truncation in the long
distance physics and  we include only the low lying charmed intermediate states. %
In order to compute eq. (\ref{eq:1}), we write
\begin{equation}
<K^0\pi^+|T\left\{J_{\mu}(x_0)\hat
  J^{\mu}(0)\right\}|B^+>\,=\,\int\frac{d^4 q}{(2\pi)^4}
  \, e^{-iqx_0}T(q)\, ,\label{eq:2}
\end{equation}
where
\be T(q) ~=~\,\int \,d^{4}x  \, e^{+iq\cdot x} <K^0\pi^+|{\rm T}
\left\{J^{\mu}(x)\hat J_{\mu}(0)\right\}|B^+> \label{Tmn} ~. \ee
Let us now show that, after averaging over $\vec n$, one obtains a
cutoff over the high frequencies in eq. (\ref{eq:2}). As a matter of
fact, one has
\begin{equation}
  A_{ChP}={G_F \over \sqrt{2}}\, a_2\,V_{cb}^* V_{cs}\,\int\frac{d^4 q}{(2\pi)^4}
\,T(q)\, \int\frac{d\vec n}{4\pi}  \, e^{-iq\cdot x_0}\,=\,{G_F
\over \sqrt{2}} a_2\,V_{cb}^* V_{cs}\,\int\frac{d^4 q}{(2\pi)^4}
\,T(q)\, \theta(\mu,\, |\vec q|)\ , \label{eq:3}
\end{equation}
where the cutoff function is
\begin{equation}
\theta(\mu,\,|\vec q|)\,=\,\frac{\sin|\vec q|/\mu}{|\vec q|/\mu}\
.\ \label{eq:4}
\end{equation}
For $\mu\to\infty$, $ \theta(\mu,\,|\vec q|)\, \to\, 1$; for
finite values of $\mu$, this function cuts off from the
$q-$integral in eq. (\ref{eq:3}) the region $-q^2\geq\mu^2$. Instead
of the smooth oscillating function (\ref{eq:4}), we used in ${\cal
I }$ the step function
\begin{equation}
  \theta(q^2+\mu^2)\ ,
\end{equation}
which allows a considerable reduction of computing
time. We also stress that one can extract from the integration in
the momentum $q$ the heavy mass contribution according to the
formula
\be q=p_B - p_{D^{(*)}}\equiv (m_B-m_{D^{(*)}}) v -\ell \ee
(see Eq.(30) of ${\cal I}$). Here $v^\mu$ is the heavy meson
velocity and $\ell$ is a residual momentum. By this the cutoff
function on the $\ell-$integration becomes
\begin{equation}
  \theta(\ell^2+\mu_\ell^2)\ ,
\end{equation}
and the value of the  cut-off $\mu_\ell$ found in ${\cal I }$ is
$0.5-0.7$ GeV. The whole procedure we have described so far has
been used several times in the past in the application of the
light cone expansion ideas to  the nonleptonic weak decays,
starting from the pioneering work of K. Wilson (see Section 7 of
\cite{wilson} and the subsequent work of several authors
\cite{altri}). We repeat it here as it may not be familiar to some
readers and also to stress that the correct value of the cut-off
$\mu$ (corresponding to $\mu_\ell$) is not the $W$ mass, but a
scale of the order of $m_b$ or, better, $m_b-m_c$. It is a
consequence of the precocity of the scaling behaviour, a
well-known example of which is provided by deep inelastic
scattering\footnote{On the basis of the previous remarks, the
criticism of ${\cal I }$ contained in \cite{beneke2} appears to be
unjustified.}. 
The introduction of non-locality through the cutoff $\mu$ of the
order of $m_b$ (or $\mu_\ell$ of the order of 0.5-0.7 GeV)  has
a clear physical meaning. It corresponds to a separation between
short distance physics (whose physical features are embodied
in the Wilson coefficients) and long-distance physics,
which is dominated by hadronic states and resonances.
In considering the long distance part, we have included the low-lying
states that could contribute, i.e., $D^{(*)}D^{(*)}_s$.
This procedure can be avoided for operators
having factorizable contributions; in the case of the charming
penguin contribution, however, such contributions do not exist, while the non
factorizable contribution are Cabibbo enhanced. This is the reason for
taking them into account explicitly.%

\noindent The second point to be stressed is that while in the present
paper, as well as in ${\cal I}$, we are using the chiral
lagrangian effective theory for heavy mesons
\cite{wise}-\cite{casalbuoni}, the light pseudoscalar mesons in
the final state have large momenta. Therefore, the effective
lagrangian must be corrected to take into account the hard  meson
momenta. The procedure we adopted in ${\cal I }$ was based on the
introduction of form factors, similar to the approach followed
in \cite{casalbuoni}, \cite{Deandrea1} and \cite{Deandrea2}. We
were able to estimate, by using the constituent quark model, the
form factor correcting the $B^*B\pi$ and $D^*D\pi$ vertices.
In addition, one should also consider the form factor correcting the
diagram in fig. 1, which represents the main contribution to the
real part of $A_{ChP}$ (in ${\cal I }$ this diagram is depicted in
fig. 2a).
\begin{figure}[ht!]
\begin{center}
\epsfig{file=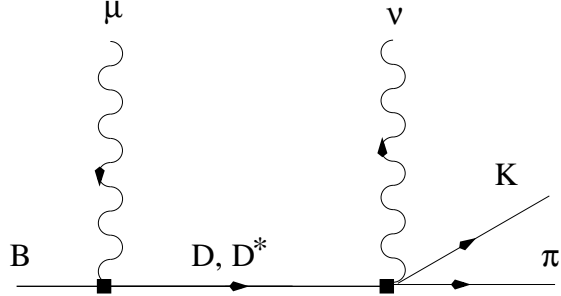,height=4cm}
\end{center}
\caption{{\small The charming penguin contribution diagram dominating the real
part of the amplitude. The boxes represent weak couplings. The $D
K \pi$ and the $D^*K\pi$ couplings represent the direct coupling
of the effective chiral theory for heavy mesons.}}
\end{figure}
The weak vertices $D^{(*)}\to K\pi$ correspond to direct,
non-resonant couplings and  arise from the weak effective current:
\begin{equation}
L_{\mu a}~=~\frac{i \alpha}{2}\ Tr \gamma _\mu (\ 1 - \gamma _5\ )
H_b
 \xi ^\dag_{ba} \; ,
\label{weak-current}
\end{equation}which is the effective realization of the quark current
${\bar q}_a\gamma^\mu(1-\gamma_5)Q$.
 $\alpha$ is related to the heavy meson leptonic decay constant
by the formula $\alpha=$ $f_D\ \sqrt{m_D}$, valid in the infinite
quark mass limit. Moreover,
\begin{equation}
H_a ~=~ \frac{1+\spur{v}}{2}\ \left( P^*_{a \mu} \gamma ^\mu - P_a
  \gamma _5 \right)
\end{equation}
and
\begin{equation}
\xi ~=~ e ^{i \frac{{\cal M}}{f}} \ .
\end{equation}
In these formulae, $v$ is the heavy meson velocity; $P_a,P^*_{a
\mu}$ are the annihilation operators of heavy pseudoscalar and
vector mesons made up by a heavy quark and a light antiquark of
flavour $a$ ($a~=~1,2,3$ for $u,~d,~s$); ${\cal M}$ is the usual
$3 \times 3$ matrix comprising the octet of pseudo--Goldstone
bosons; and $f\simeq f_\pi\approx 132\,\, MeV $ is the
pseudo--Goldstone bosons decay constant. Equation (\ref{weak-current})
generates not only weak couplings of $D,~D^*$ to hadronic final
states with two pseudo--Goldstone bosons, but also the amplitudes
with one light pseudoscalar boson in the final state; in
particular, it produces the Callan-Treiman relation relating the
form factor $F_0^{D\pi}(q^2)$ at $q^2=m_D^2$ with $f_D$, {\it
i.e.},
\begin{equation}
F_0^{D\pi}(m_D^2)\simeq\frac{f_D}{f_\pi}\ .
\end{equation}
At the scale we are interested in, this vertex should be corrected
by a form factor that we call $F_a(q^2)$ ($q$ is the four momentum
carried by the current),
\begin{equation}
\frac{f_D}{f_\pi}\ \to \ \frac{f_D}{f_\pi}\, F_a(q^2)\ .
\end{equation}
We do not have sufficient information on the behaviour of
$F_a(q^2)$\footnote{A model calculation of this form factor is in
\cite{ladisa2}.}, therefore, we leave it as a (constant) parameter
and we write
\be F_a=1.0\pm0.5\ . \label{19} \ee
It must be stressed, however, that, in the evaluation of the scaling
behaviour (with $1/m_b$) of the charming penguin contributions, the role of
this form factor is indeed relevant. Assuming \footnote{This
assumption is based on the dominance of the hard contribution in
the QCD evaluation of the form factor; the actual scaling law may
be affected by the behaviour in the soft region, {\it e.g.} at the
end points.}, as in \cite{beneke}, that $F^{B\pi}_0(m^2_K)$ scales
as $\displaystyle\left(\frac{\Lambda}{m_b}\right)^{\frac 3 2}$, one
gets  a scaling law $m_b^{\frac 1 2}$ for the factorized
contribution (in our language $A_{T+P}$). However, in these
hypotheses, the form factor $F^{B\pi}_0(q^2)$ should display a
pole behaviour, to match with the
$\displaystyle\left(\frac{\Lambda}{m_b}\right)^{\frac 1 2}$
behaviour predicted by the Callan-Treiman relation at $q^2\simeq
m_B^2$. One can now assume  a similar behaviour for the form
factor $F_a$, which is reasonable, as the two amplitudes $P\to M$
and $P\to MM$ ($P$ heavy, $M$ light mesons) derive from the same
effective current $L^\mu$.  This implies that the contribution of
the charming penguin contribution diagrams should be suppressed by some power
${\cal O} (1/m_b)$ in the $m_b\to\infty$ limit in comparison with
the factorizable ones. As we are not able to define better the
form factor $F_a$, our evaluation should be understood as an order
of magnitude  estimate.

\section{Results}
\noindent Given these remarks, we are now ready to present our
results. The Tree and Penguin contribution to the decay processes
$B\to K\pi$, $B\to\pi\pi$, and $B\to K\eta^{(\prime)}$ is obtained
by the usual procedure of factorization using the non leptonic
hamiltonian as given {\it e.g.} in \cite{Ali}. As  for the
charming penguin contribution terms, the explicit formulae can be found in
${\cal I }$ and need not be reported here (in ${\cal I }$ they
are denoted as $A_{LD}$). The numerical results we obtain for the
amplitudes are reported in Table 1.
\begin{table}[ht!]
\caption{{\small  Theoretical values for $A_{T+P}$ (Tree+Penguin
amplitude) and $A_{ChP}$ (Charming Penguin amplitude).}}
\begin{center}
\begin{tabular}{|c|c|c|}
\hline {\rm Process} & $A_{T+P}\times 10^8$ GeV  &  $A_{ChP}\times 10^8$ GeV  \\
\hline
$  {B}^{+}\, \to \,K^{0}\pi^{+}$ & $+1.69$ & $+2.06\ +\ 2.36\,i$\\
\hline ${B}^{+}\, \to \,K^{+}\pi^{0}$ &
$+1.21\ -\ 0.498\,i$  & $ +1.45\ +\ 1.67\,i $ \\
\hline ${B}^{0}\, \to \, K^{+}\pi^{-}$ &
$+1.32\ -\ 0.634\,i$  & $+2.06\ +\ 2.36\,i$ \\
\hline ${B}^{0}\, \to \,K^{0}\pi^{0}$ &
$-0.921\ -\ 0.0497\,i$  & $-1.45\ -\ 1.67\,i $ \\
\hline $ {B}^{+}\, \to \, \pi^{+}\pi^{0}$&
$-1.35\ -\ 1.79\,i $  & $ 0 $\\
\hline ${B}^0\, \to\, \pi^{+}\pi^{-}$ &
$-1.85\ -\ 2.16 \,i$  & $ -0.576\ -\ 0.648\,i $\\
\hline ${B}^{0}\, \to \,\pi^{0}\pi^{0}$&
$+0.0516\ +\ 0.379 \,i$& $ -0.576\ -\ 0.648\,i$\\
\hline $ {B}^{+} \,\to\, K^{+} \eta$ &
$-0.0491\ -\ 0.415 \,i$  & $ +0.0830\ + \ 0.0896\,i  $\\
\hline $ {B}^{+} \,\to\, K^{+}\eta^\prime$ &
$+1.40\ -\ 0.261 \,i$ & $ +2.53\ +\ 2.83\,i  $\\
\hline ${B}^{0} \,\to\, K^{0}\eta$ &
$~ +0.172\ -\ 0.0418 \,i$  & $ +0.0830\ +\ 0.0896\,i$\\
\hline ${B}^{0} \,\to\, K^{0}\eta^\prime$ &
$+1.54 \ -\ 0.0269\,i $ & $+2.53\ +\ 2.83\,i $\\
\hline
\end{tabular}
\end{center}
\end{table}
We note that the phase of $A_{T+P}$ is due only to the weak
interactions, while the phase in $A_{ChP}$ is purely strong. We
use the following set of parameters (with the notations of
 \cite{Ali}):  For the Wilson coefficients \cite{Buras}:
$c_2=1.105,~c_1=-0.228,~c_3=0.013,~c_4=-0.029,~c_5=0.009,
~c_6=-0.033,~c_7/\alpha=0.005,~c_8/\alpha=0.060,
~c_9/\alpha=-1.283$, and $c_{10}/\alpha=0.266$. Moreover, we use
$F_0^{B\, M'}(m^2_M)\approx F_0^{B\, M'}(0)=0.25$
($M,~M'=K,~\pi^\pm$)\footnote{We employ the QCD Sum Rule result of
\cite{santorelli}; a slightly higher value is in
 \cite{ball}.}.
 The amplitudes are evaluated using, for the CKM matrix
elements, the results of the  analysis \cite{ciuchini}:
$A=0.82,~\rho=0.23$, and $\eta=0.32$. For the $K\eta^{(\prime)}$ final
state, we use $SU(3)$ symmetry and the method of \cite{Ali} with
$f_0=f_8=f_\pi=132$ MeV and $\theta_0=\theta_8=-22^o$. We also
notice that our phase convention is such that the amplitude ${\cal
A}(B^+\to K^0\pi^+)$ differs by a sign from the result of
\cite{neubert}; for $B\to\pi^0\pi^0$, the statistical factor 1/2 in
the branching ratio takes into account the identity of the final
mesons.
 \begin{table}[ht!]
\caption{{\small  Theoretical values for the  CP averaged
Branching Ratios (BR) compared with experimental data. Data are
averages \cite{martinafranca} from  among CLEO \cite{cleoprl2000},
BaBar \cite{babarmarch2001}, Belle \cite{Bellefeb2001} except for
the upper limit that comes from \cite{cleoprl2000}. }}
\begin{center}
\begin{tabular}{|c|c|c|c|}
\hline {\rm Process}  & {\rm BR $\times 10^{6}$ (T+P)}& {\rm BR
$\times 10^{6}$ (T+P+ChP)} &
 {\rm BR $\times 10^{6}$ (Exp.)} \\
\hline
${B}^{\pm}\, \to \,K^{0}\pi^{\pm}$ &$\sim 2.7\,\, $& $ 18.4\ \pm\ 10.8 $ & $ 17.2\,\pm\,2.5$\\
\hline
${B}^{\pm}\, \to \,K^{\pm}\pi^{0}$ &$\sim 1.6\,\, $& $ 9.5\ \pm\ 5.5  $ & $ 12.1\,\pm\,1.7 $\\
\hline
${B}\, \to \, K^{\pm}\pi^{\mp}$ &$\sim 1.9\,\, $& $ 15.3\ \pm\ 9.9 $ & $ 17.2\,\pm\,1.5$\\
\hline
${B}^0\, \to \,K^{0}\pi^{0}$ &$\sim 0.75 $& $ 7.4\ \pm\ 4.8 $ & $ 10.3\,\pm\,2.5$\\
\hline
${B}^{\pm}\, \to \, \pi^{\pm}\pi^{0}$ &$\sim 4.8\,\, $& $  \sim 4.8 $ & $ 5.6\,\pm\,1.5$\\
\hline
${B}^0\, \to\, \pi^{+}\pi^{-}$ &$\sim 7.2\,\, $& $ 9.7\ \pm\ 2.3 $ & $4.4\,\pm\,0.9$\\
\hline
${B}^0\, \to \,\pi^{0}\pi^{0}$&$\sim 0.06 $& $ 0.37\ \pm \ 0.35 $ & $ < 5.7$ \\
\hline
\end{tabular}
\end{center}
\end{table}
From the results in Table 1, we can compute the Branching Ratios
(BR) and the CP asymmetries for the $K\pi$ and $\pi\pi$ final
states. The CP averaged Branching Ratios are reported in Table 2.
In the first numerical column, we report the results obtained by
including only the Tree and Penguin contributions, {\it i.e.},
$A_{T+P}$; in the second column, we give the results obtained by
the full amplitude $A_{T+P}+A_{ChP}$; in the final column,
we give the available data
from the CLEO, Belle and BaBar experiments. The errors on the
branching ratios are obtained varying independently the cut-off
$\mu_{\ell}$ in the range $0.5 \div 0.7$ GeV, $F_a$ in the range
$0.5\div 1.5$, and $F( |{\vec p_{\pi}| })\ =\ 0.065 \pm 0.035$,
and summing the errors in quadrature. We have not added the errors
related with the Tree and Penguin contribution, arising from the
CKM matrix elements and from the hadronic parameters. A comparison
between and the first and the second column shows the importance
of $A_{ChP}$ for the $K\pi$ final state, while for the $\pi\pi$
final states the charming penguin contribution is either absent
($\pi^\pm\pi^0$) or less important ($\pi^+\pi^-$). As a matter of
fact, as already observed in ${\cal I}$, using $SU(3)$ symmetry
one obtains for this channel
\be A_{\rm ChP}({B}^{0} \to \pi^{+}\pi^{-}) =
\frac{V_{cd}}{V_{cs}}\,\,\frac{f_K}{f_\pi}\,\, A_{\rm ChP}({B}^{0}
\to K^{+}\pi^{-}) ~, \label{Bpipi} \ee
{\it i.e.}, a CKM suppression in comparison with the $K\pi$ final
state.
We note a general
good agreement with the data; the only significant difference is
for the $\pi^+\pi^-$ final state, which in our opinion should be
explained by a more refined analysis of the errors in the inputs
of the Tree and Penguin contributions (we repeat that, for the
sake of simplicity, we have not introduced these errors in our
discussion). In any event, Table 2 show that the uncertainties
arising from the charming penguin contribution term are rather large.

The absorptive part of $A_{ChP}$, which is less sensitive to
theoretical uncertainties than the real part, provides a strong
argument for a large inelastic final state interaction phase and
for an appreciable CP violation even in the absence of the $ K\pi$
and $\pi\pi$ elastic rescattering phase shift. To be more
quantitative, from the results in Table 1 we compute $CP$
violating asymmetries for the various channels:
\begin{eqnarray}
{\cal A}^{+-}_{\pi\pi}&=&\frac{ BR( \overline{B^0} \to
\pi^+\pi^-)-BR(B^0\to \pi^+\pi^-)} {BR(\overline{B^0}\to
\pi^+\pi^-)+BR(B^0\to \pi^+\pi^-)}\ ,\cr && \cr
{\cal A}^{-0}&=& \frac{ BR(B^-\to K^-\pi^0)-BR(B^+\to K^+\pi^0)}
{BR(B^-\to K^-\pi^0)+BR(B^+\to K^+\pi^0)} \ , \cr & & \cr
{\cal A}^{0-}&=& \frac{ BR(B^-\to \overline{K^0}\pi^-) - BR(B^+\to
K^0\pi^+)} {BR(B^-\to \overline{K^0}\pi^-)+BR(B^+\to K^0\pi^+)}\ ,
\cr & & \cr
{\cal A}^{-+}&=& \frac{ BR(\overline{B^0}\to K^-\pi^+)-BR(B^0\to
K^+\pi^-)} {BR(\overline{B^0}\to K^-\pi^+) + BR(B^0\to K^+\pi^-)}\
, \cr & & \cr
{\cal A}^{00}&=& \frac{BR(\overline{B^0}\to \overline{K^0}\pi^0) -
BR(B^0\to K^0\pi^0)} {BR(\overline{B^0}\to \overline{K^0}\pi^0) +
BR(B^0\to K^0\pi^0)}\ .
\end{eqnarray}
We obtain the following results:
\begin{equation}
{\cal A}^{+-}_{\pi\pi} = \frac{ BR( \overline{B^0} \to
\pi^+\pi^-)-BR(B^0\to \pi^+\pi^-)} {BR(\overline{B^0}\to
\pi^+\pi^-)+BR(B^0\to \pi^+\pi^-)}\ \ =\ -\,0.24\,\pm ~0.24\
,\label{asipipi}
\end{equation}
while the  asymmetries  for the $K\pi$ final state are reported in
Fig. 2 as a function of the angle $\gamma=arg(\, V_{ub}^*\,)$. We
have not reported the asymmetry ${\cal A}^{0-}$ that vanishes in
our approach. The regions reported in these graphs correspond to a
variation of the three most relevant parameters affecting our
numerical results, {\it i.e.}, the cutoff $\mu_\ell~\in~[0.5,~0.7]$
GeV and the form factors $F_a$ in eq. (\ref{19}) and $F(|\vec p_\pi|)\
\in\ [\,0.03,\,0.10\,]$. We see that the variations are rather
large, but still compatible with the CLEO \cite{CLEOCP}, BaBar
\cite{BBCP}, and Belle data \cite{BelleCP}, which are as follows
\cite{martinafranca}:
\begin{equation}
{\cal A}^{-0}\, =\, -0.096\, \pm\, 0.119\, , \hspace{0.5truecm}
{\cal A}^{-+}\, =\, -0.048\, \pm\, 0.068\, , \hspace{0.5truecm}
{\cal A}^{0-}\, =\, -0.047\, \pm\, 0.139\, .
\end{equation}
\begin{figure}[ht!]
\begin{center}\begin{tabular}{lcr}
&\epsfig{file=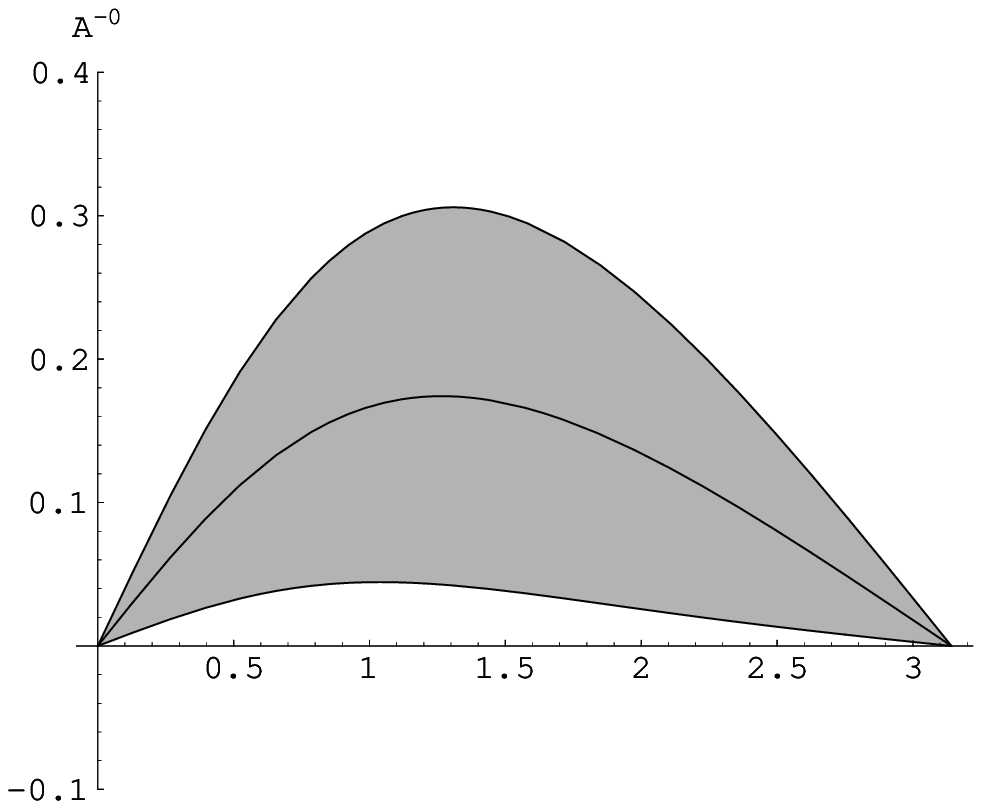,height=4.0cm}\hskip2cm
\epsfig{file=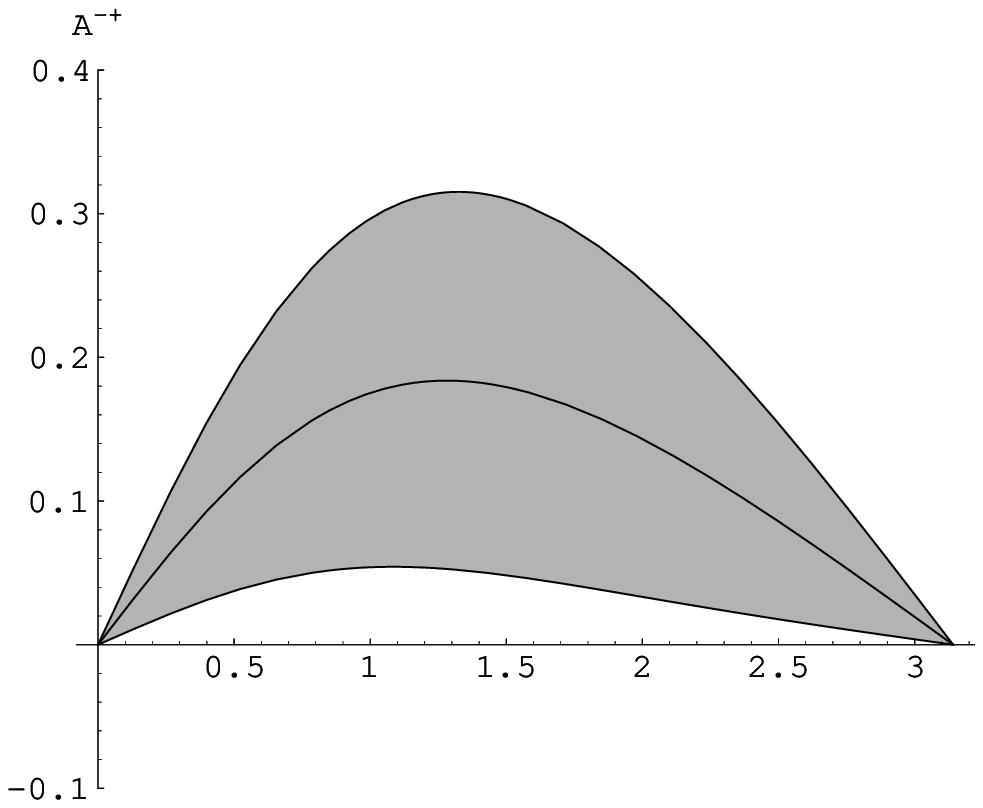,height=4.0cm}&\\
&\epsfig{file=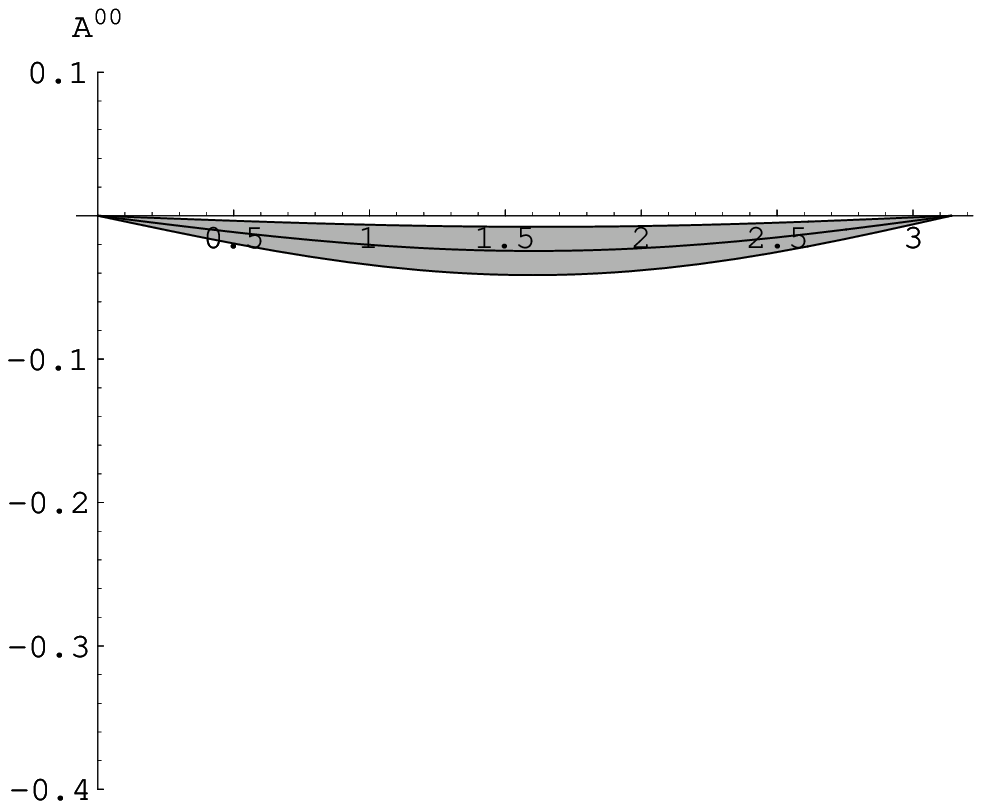,height=4.0cm}&
\\
\end{tabular}
\end{center}
\caption{{\small First line, CP asymmetries ${\cal A}^{-0}$ (left)
and ${\cal A}^{-+}$ (right); second line ${\cal A}^{00}$. The
asymmetries are plotted versus the angle $\gamma$ .}}
\end{figure}
The CP asymmetries we obtain are large, about 20\% or more, as
shown in Fig. 2. In particular, we find large CP asymmetries for
$B^{0}\to \pi^+\pi^{-}$ decays, see Eq. (\ref{asipipi}). The
measurement of the weak angle $\alpha$ from $B^{0}\to \pi^+\pi^{-}$ decays could
still be possible once an accurate determination of the
long-distance absorptive part from $B\to K\pi$ decays was
obtained. Our results for the asymmetries are (in absolute value)
compatible with Refs. \cite{strikeback}, \cite{ciuchini2}, which
also obtain  large CP asymmetries for $B \to K\pi,\pi\pi$ decays
in  phenomenological analyses of the charming penguin
contributions. This is in contrast with the QCD-improved
factorization model, which predicts small CP asymmetries for
$B\to K\pi$ and $B\to \pi\pi$ decays
\cite{beneke,martinafranca,muta}.

Let us also briefly comment on the $K\eta $ and $K\eta^\prime$
final states. Our results for these channels are reported in Table
3. For the $K\eta^\prime$ final states, one can clearly see that
the charming penguin contribution significantly enhances the results and may
be important for producing a large branching ratio; however, it is
also clear that some relevant further contribution is still
missing since by no reasonable choice of the parameters can the
charming penguin contribution alone solve the puzzle posed by
experimentally very large decay fractions. We refer the reader to
the existing literature \cite{cinesi} on this subject.
\begin{table}[ht!]
\caption{{\small  Theoretical values of Branching Ratios (BR) for
$B\to K\eta^{(\prime)}$ compared with experimental data from (a)
CLEO \cite{cleoprl2000}; (b) average between Belle
\cite{Bellefeb2001} and CLEO \cite{cleoprl2000}. }}
\begin{center}
\begin{tabular}{|c|c|c|c|}
\hline
{\rm Process} &  {\rm BR $\times 10^{6}$ (T+P)} & {\rm BR$\times 10^{6} $ (Ch+T+P)} & {\rm BR$\times 10^{6}$ (Exp.)}\\
\hline $  {B}^{+} \,\to\, K^{+}\eta$ & $\sim 0.162$ & $0.099\ \pm\ 0.029 $ & $ -- $  \\
\hline $  {B}^{+} \,\to\, K^{+}\eta^\prime$ & $\sim 1.83$ &$ 20\ \pm\ 10  $ & $ 80.0\,\pm\,12.2\ $(a)\\
\hline ${B}^0 \,\to\, K^{0}\eta$ & $\sim 0.027$ &$ 0.058\ \pm\ 0.016 $ & $ --$ \\
\hline ${B}^{0} \,\to\, K^{0}\eta^\prime$ & $\sim 2.00$ &$ 21\ \pm\ 10 $  & $ 80\,\pm\,15\ $(b)\\
\hline
\end{tabular}
\end{center}
\end{table}

\section{Conclusions}

In conclusion, we have extended our model of the charming penguin
contributions in $B\to K\pi$ decays to all the significant decays
with two pseudoscalar mesons in the final state. Although the
calculation presents a number of theoretical uncertainties, it
clearly shows that the effect of the charming penguin contribution terms is
overwhelming for all the $B \to K\pi$ decay modes while its role
is less significant in the other channels. This dominance is not
parametric, {\it i.e.}, it does not contradict the  dominance of
the factorized amplitude in the $m_b\to\infty$ limit discussed by
several authors in the last two years \cite{beneke2},
\cite{beneke}, \cite{du}, \cite{keum}, \cite{mehrban}. It arises
from the  CKM enhancement of the non-factorized decay chains
(\ref{cp}) and their related real parts. The size of these
charming penguin contribution terms can be estimated by an effective field
approach, though a complete calculation is beyond the presently
available theoretical methods. Therefore, one cannot escape the
conclusion that, in spite of the proven theorems, the elusive
non-leptonic B-decays still maintain their secrecy.
\section*{Acknowledgements}
M.L. acknowledges partial support from the Israel-USA Binational
Foundation and from the Israel Science Foundation.

\end{document}